\documentclass[aps,pra,preprint,groupedaddress,showpacs]{revtex4-1}

\usepackage[latin1]{inputenc}
\usepackage{amsmath}
\usepackage{amsfonts}
\usepackage{amssymb}
\usepackage[dvips]{graphicx,graphics}
\usepackage{epsfig}
\usepackage{xcolor,soul}

\usepackage{ulem} 

\sethlcolor{yellow}
\setulcolor{red}
\newcommand{\beq}{\begin{equation}}
\newcommand{\eeq}{\end{equation}}

\begin{document}
\title{Ion transport in macroscopic RF linear traps}
\author{Jofre Pedregosa-Gutierrez}
\email[]{jofre.pedregosa@univ-amu.fr}
\author{Caroline Champenois}
\author{Marius Romuald Kamsap}
\author{Martina Knoop}
\affiliation{ Aix-Marseille Universit\'e, CNRS, PIIM, UMR 7345, 
 Centre de Saint J\'er\^ome,
13397 Marseille Cedex 20, France}

\date{\today}

\begin{abstract}
Efficient transport of cold atoms or ions is a subject of increasing concern in many experimental applications reaching from quantum information processing to frequency metrology. For the scalable quantum computer architectures based on the shuttling of individual ions, different transport schemes have been developed, which allow to move single atoms minimizing their energy gain. In this article we discuss the experimental implementation of the transport of a three-dimensional ion cloud in a macroscopic linear radiofrequency (RF) trap. The present work is based on numerical simulations done by molecular dynamics taking into account a realistic experimental environment. The deformation of the trapping potential and the spatial extension of the cloud during transport appears to be the major source of the ion energy gain. The efficiency of transport in terms of transfer probability and ion number is also discussed.\end{abstract}

\pacs{37.10.Ty (Ion trapping) 37.10.Rs (Ion cooling) 06.30.Ft (Time and frequency ) }

\maketitle

\section{Introduction}
Since the very first experiments of guided ion beams by radiofrequency (RF) electric fields \cite{teloy74}, the transport of ions is a relevant issue in experiments involving trapped ions. For many different applications it is important to shuttle ions from one trap to another, the main conditions being the absence of heating of the atoms and the maximisation of the transported ion number. Ion transport has become a standard item in mass spectrometry experiments, where it is often coupled to external ion creation, by various techniques reaching from Electrospray Ionization (ESI) (see for example \cite{senko97}) to the production of exotic ions at CERN \cite{herfurth03}. In most of these experiments, the efficiency of transport has to be high, without necessarily reaching 100 \%. In some situations, the use of buffer gas increases performances as it damps eventual heating. Different geometries of linear RF guides can be used, from the most popular quadrupolar geometry to the 22-pole one. Already in the early years, ion transport was shown to depend on the RF-phase of the trapping field \cite{haegg86}.

Recent advances in quantum information processing, in particular the different realizations of scalable architectures \cite{kielpinski02} rely on the transport of single ions. Transported ions have to be kept in a very low vibrational level, if not the ground vibrational state, with extremely high fidelity. Very low ion temperatures are reached by laser-cooling, without the need for a thermalizing buffer gas. These experiments require the shuttling of the ion, which means that the sample is transferred many times back and forth between the different parts of the trapping device. The duration of the transport process is an issue, as it constitutes a dead-time between two gate operations and the challenge consists in reaching the low heating constraint outside the adiabatic regime. Two recent experiments demonstrated fast transport of a single ion with an energy gain as low as 0.1 vibrational quanta : Walther {\it et. al} report the shuttling of an ion over 280~$\mu$m in 3.6~$\mu$s which represents five oscillation periods \cite{walther12} whereas Bowler {\it et. al} shuttled their ion over 370~$\mu$m in 8~$\mu$s which represents 16 oscillation periods \cite{bowler12}. 

Concerning many-body systems, transport of an ensemble of neutral atoms has been demonstrated for cold atoms and Bose-Einstein condensates (BECs) making use of shortcuts to adiabacity \cite{couvert08,schaff11a}. In \cite{couvert08} a cloud of a few $10^6$ cold atoms is shuttled back and forth in an optical tweezer by a distance of 22.5 mm, using times as short as 4 oscillation periods. The use of an optical tweezer is very advantageous as it can be moved with minimal deformation. The faster than adiabatic transport scheme used in \cite{couvert08} relies on this non-deformation and it can not be extrapolated to trapping potentials in large RF traps, as justified in \ref{s:implementation}. The scheme designed in \cite{schaff11a} for the decompression evolution was engineered using dynamics invariants and the scaling of the equation of motion of an non-interacting gas and of a BEC in the Thomas-Fermi limit. Such decompression generates a vertical displacement of the centre of mass due to the competition between the gravity and the trapping potential. The long-range Coulomb repulsion makes it difficult to adapt this shortcut to adiabaticity scheme to a set of trapped ions.

In this article we discuss the experimental implementation of the transport of a three-dimensional ion cloud in a macroscopic linear radiofrequency (RF) trap. Our experiment \cite{pedregosa10a,champenois13} is designed to investigate the dynamics and thermodynamics of large ion clouds in potentials of different geometry. The double trap which combines a quadrupole and an octupole confinement zone is similar to the trap developed at the Jet Propulsion Laboratory (JET) \cite{prestage03}. This type of configuration requires the shuttling of the trapped ions back and forth between the different trapping zones, as was first demonstrated by JD Prestage et al. \cite{prestage95}. The control of the dynamics of very large clouds is also interesting for the study of frequency standards in the microwave regions, where wavelengths in the $cm$-range assure trapping in the Lamb-Dicke domain \cite{dicke53} for all millimeter-sized traps.

Experimentally, we are able to shuttle cold ion clouds in our double trap with an efficiency higher than 90\%~\cite{kamsap14}. In order to control heating processes during transport, we rely on numerical simulations based on molecular dynamics. The code realistically describes the ions in their environment, by taking into account all relevant forces (trapping potential, Coulomb interaction) as well as laser-cooling as a stochastic process. The mechanical effect of light is implemented by momentum kicks induced by photon absorption and emission, like described in \cite{bluemel88,marciante10}. Trap potentials can be described either analytically or by directly making use of the SIMION potential grid reproducing the experimental geometry of the electrodes. Throughout this paper, if not otherwise indicated, we use the standard experimental parameters of our set-up within the quadrupole part\cite{champenois13}, as well as the calcium ion mass for the molecular dynamics simulations.

The present manuscript is organized as follows. The following section is devoted to the energy gain of a single ion transported along a 1D translated potential. Different gates are compared and the relevant time scales are identified. The analogies between the single ion energy gain and the same figure of merit for the centre of mass (CM) of an ion cloud are then shown in the case of an ideal translated harmonic potential. In section \ref{s:implementation}, we introduce the constraints and limits induced by the experimental implementation of a given gate in a macroscopic trap designed for large clouds. We focus on the energy gain due to the time variation of the harmonic potential and to the non-harmonic contribution of the potential. In the last subsection, the effect of the number of trapped ions on the transport efficiency is analyzed.

\section{Ion transport by translating a harmonic potential}\label{s:w0}
\subsection{Single ion shuttling}\label{s:1D_w0}
In the limit case where ions are trapped in a harmonic potential of constant steepness, the CM of an ion cloud follows the same dynamics as a single ion \cite{palmero13}. Therefore, we use the single ion case as a starting point for our study of an ion cloud transport. In the ideal case where the ion can be transported by translating a stationary harmonic potential characterized by a constant eigen-frequency $\omega_0$ and a moving minimum, $z_{min}(t)$, the trapping potential can be written as
\begin{equation}\label{eq:ideal_potential_w0}
U_0(t) = \frac{1}{2} m \omega_0^2 ( z - z_{min}(t) )^2
\end{equation}
where $m$ is the mass of the ion. 


Assuming an ion initially at rest at the equilibrium position, the final energy of the ion after a transport of duration $t_{gate}$, can be obtained analytically~\cite{reichle06} by introducing a generalised kinetic energy $E(t_{gate}) = m|\Xi(t_{gate})|^2 /2$ which depends on the acceleration of the potential minimum along the transport like
\begin{eqnarray}\label{eq:Xi}
|\Xi(t_{gate})|^2 = \left(\int_{0}^{t_{gate}}{\cos(\omega_0 t)\ddot{z}_{min}(t)dt}\right)^2 \nonumber \\
 + \left(\int_{0}^{t_{gate}}{\sin(\omega_0 t)\ddot{z}_{min}(t)dt}\right)^2.
\end{eqnarray}
This expression is a well known result connecting the transferred energy to the Fourier transform of the force pushing the ion, at frequency $\omega_0$. It can serve as a figure of merit to compare different gates, like done in \cite{reichle06, hucul08} where three different kinds of potential-minimum time profiles are compared, which stand for different characteristic behavior of the initial, final, and average acceleration. Since the publication of this comparison, Torrontegui {\it et al.} designed a polynomial gate for fast transport without heating based on the dynamical invariants associated to the Hamiltonian \cite{torrontegui11}.

In this paper, we compare the above mentioned four analytical potential-minimum time profiles, for the single ion case and in the ion cloud regime. The general expression for these time profiles is
\begin{equation}
z_{min}(t) = f_{i}(t)( H(t) - H(t-t_{gate}) ) + L H(t-t_{gate})
\end{equation}
with $H(t)$ the Heaviside step function and $f_i$, one of the analytic gates listed below, with $s=t/t_{gate}$ the relative gate duration and $L$ the shuttling distance :
\begin{eqnarray}
f_{lin}(s) &= Ls \label{eq:lin} \\
f_{sin}(s) &= \frac{L}{2}\left(1 - \cos\left(\pi s \right) \right) \label{eq:sin} \\
f_{tanh}(s) &= \frac{L}{2}\left(\frac{ \tanh\left( 2N_Hs - N_H \right)}{\tanh(N_H)} + 1\right) \label{eq:tanh}\\
f_{poly}(s) &= L \Big( \frac{60s-180s^2+120s^3}{t^2_{gate}\omega^2_0} \nonumber\\
&\quad+ 10s^3 -15s^4 + 6^5\Big) \label{eq:poly}
\end{eqnarray}
The impact of the $N_H$ coefficient of Eq.(\ref{eq:tanh}) is analyzed in \cite{hucul08} and we choose $N_H=4$ throughout this paper as a compromise between a linear (N = 1) and a step function ($N \to \infty$). The polynomial gate designed in \cite{torrontegui11} is the only one which depends explicitly on the considered harmonic potential. The final energy of a single ion shuttled by each of these gates can be analytically derived. The expressions for the expected energy for the linear, sinusoidal and polynomial profiles as a function of the oscillation phase during the shuttling $\theta=\omega_0 t_{gate}$ are:

\begin{eqnarray}
|\Xi_{lin}(t_{gate})|^2&=& 4L^2\omega_0^2\left(\frac{\sin\left(\theta/2\right)}{\theta}\right)^2 \label{eq:Xi_lin}\\
|\Xi_{sin}(t_{gate})|^2 &=&\pi^2 L^2\omega_0^2 \left(\frac{\cos\left(\theta/2\right)}{\pi^2 - \theta^2} \right)^2 \label{eq:Xi_sin}\\
|\Xi_{poly}(t_{gate})|^2 &=& 14400 L^2\omega_0^2\left( \frac{\sin(\theta/2)}{\theta^3 } \right)^2 \label{eq:Xi_poly}
\end{eqnarray}

\begin{figure}[htb]
\begin{center}
\includegraphics[width=10cm]{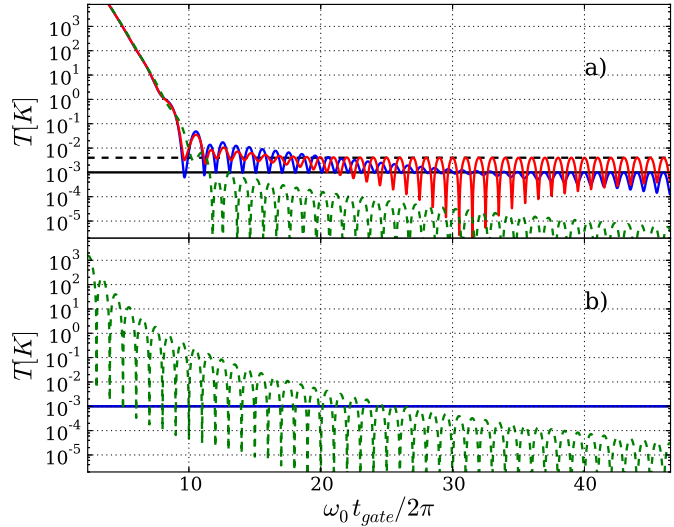}
\caption{(Color online) Total energy (converted into temperature) of a single ion after transport over a distance of 23~mm, for different transport durations in units of relative oscillation phase $\omega_0t_{gate}/2\pi$. The potential minimum obeys (a) the hyperbolic tangent time profile of Eq~(\ref{eq:tanh}) and (b) the polynomial time profile defined by Eq~(\ref{eq:poly}). Green dashed line: Eq~(\ref{eq:Xi}), Blue solid line: obtained from solving the equations of motion of an ion with an initial velocity equivalent to a temperature of 1~mK. Red solid line: as for the blue line with an initial velocity equivalent to a temperature of 4~mK. The horizontal black lines stand for the 1~mK and 4~mK initial energies. \label{fig:ideal_vs_analytic}}
\end{center}
\end{figure}

The hyperbolic tangent leads to a very complex expression that can be found in Eq. 101 of \cite{hucul08}. For comparison, we computed numerically $|\Xi(t_{gate})|^2$ for the hyperbolic tangent case. This final energy is converted into temperature and plotted on figure~\ref{fig:ideal_vs_analytic} (green lines) and compared with the polynomial case. 
The evolution with the shuttling duration share the same features : a decreasing envelope for the maximum energy gain, and a periodic cancellation of the energy gain related to the harmonic oscillation in the potential. The hyperbolic tangent profile is the only one giving rise to a behavior with two very different time scales, like noticed in \cite{hucul08} : the periodic cancellation appears only for a minimum value of $\theta$, which depends on the chosen $N_H$. For shorter shuttling times, the energy gain reaches very high values and decreases drastically as a function of duration. For longer shuttling time constants, the envelope shows a continuous decrease that is faster than that of the other three gates. On long time scales, a comparison of the expected energies show that the hyperbolic tangent gate seems to be the most advantageous for transport without heating. 

The energy gain expected from Eq.~\ref{eq:Xi} has been experimentally verified with cold atoms in an optical tweezer like described in the introduction. The periodic cancellation of the energy gain allows to realize faster than adiabatic transport with no excitation of the center of mass (CM) motion \cite{couvert08}.

As we want to extrapolate the single ion case to the transport of an ensemble of ions with different initial conditions, we start by computing the transport-induced energy gain for a single ion with a non-null initial velocity. This is done by integrating the equation of motion during the transport, including the force derived from the moving harmonic potential of Eq.(\ref{eq:ideal_potential_w0}), for an ion initially at the potential minimum with a kinetic energy corresponding to a temperature of 1mK, its velocity vector oriented towards the shuttling destination. 

In the case of the linear and sinusoidal potential-minimum time profiles, no major differences can be observed, except for the values of the periodic minima: while the linear profile results in a periodic cancellation of the energy gain, the sinusoidal gate cancels the total energy. This difference can be of importance in single ion shuttling experiment if the ion's energy needs to be reduced.

The blue lines in Figure~\ref{fig:ideal_vs_analytic} show the results of the simulation for the hyperbolic tangent and the polynomial gates which differ strongly from the immobile ion case. In the polynomial case, the final energy completely looses its periodic features and equals exactly the initial energy. Actually, this gate has been designed for an immobile start and arrival \cite{torrontegui11}, it is very robust against a non-zero initial velocity and no energy gain was observed from this gate in the velocity range we explored. On the other hand , an ion shuttled with a hyperbolic tangent gate still shows a final energy with a periodic structure when starting with non-zero initial energy. For short enough shuttling durations, the minimum energy equals the initial energy whereas after a few tens of oscillation periods, the final energy after shuttling is reduced below the initial value. The exact gate duration for which an energy loss starts to be observed depends on the initial velocity of the ion: the larger the initial velocity, the shorter shuttling it takes for an ion to reduce its energy.

\subsection{Ion cloud transport}\label{s:3D_w0}
As demonstrated in \cite{palmero13} for two ions, the motion of the CM of an ion cloud is expected to follow the dynamics of a single ion if the translated potential is harmonic with a constant oscillation frequency $\omega_0$. Extending this analogy to a cloud of $N_0$ ions with Coulomb repulsion requires to consider the dynamics in 3D. In a first step, we choose the trapping potential in an ideal RF linear quadrupole trap
\begin{eqnarray}
\label{eq:idealRF}
\Psi(t,x,y,z) &= (V_{DC}-V_{RF}\cos(\Omega t))\frac{x^2 - y^2}{r_0^2} \nonumber \\ 
& + \frac{1}{2} m \omega_0^2 \left[( z - z_{min}(t) )^2- \frac{x^2 + y^2}{2} \right]
\end{eqnarray}
We set the trap parameters in the adiabatic regime \cite{dehmelt67} with a Mathieu parameter $q_x=0.14$ ($V_{RF}=250$~V, $\Omega/2\pi=5.25$~MHz for $r_0=3.93$~mm). $V_{DC}$ is fixed at 1~V, a value too small to modify the potential depth but sufficient to prevent the cloud rotation observed in this kind of simulation in axially symmetric potentials \cite{marciante10}. In the adiabatic approximation, the radial oscillating potential is equivalent to a harmonic static potential characterized by $\omega_x/2\pi=281$~kHz. Furthermore, we choose for $\omega_0$ a value which leads to an aspect ratio $R/L$ of the ion cloud in the cold limit equal to 0.31, which corresponds to a 3D morphology with a cigar shape \cite{turner87,drewsen98}. This implies that $\omega_0/2\pi = 124$~kHz and the effective radial harmonic potential is given by $\omega_r =267$~kHz. Such values correspond to our experimental set-up as described in \cite{champenois13}.

Before shuttling, the initial conditions of each ion are prepared in three steps : first, the ions are set at random positions following a Gaussian distribution, and with zero velocity. From this moment on, they are submitted to the trapping force and the Coulomb repulsion. After 100 RF periods, where no cooling is applied, the thermal bath technique is used: at a time for which the RF-driven velocity is null, the ion velocity is periodically rescaled in the three directions such as to reach a temperature of the ensemble equal to $T_{bath}=1$ ~mK to within a 1\% precision \cite{nose84}. Finally, the thermal bath is turned off and the ions are submitted to laser Doppler cooling for an evolution time of 5~ms. 
In our case, the laser beam propagates along the trap symmetry axis, which is sufficient to cool the ion cloud in the three degrees of motion because of its 3D morphology \cite{marciante10}. We treat the atomic system as a two-level atom and chose laser interaction parameters in the slightly saturated regime (s=1.5 for a detuning corresponding to 2.5 times the natural linewidth) which allows the velocity distribution to reach an equilibrium corresponding to a temperature $T \approx 4$~mK. With the chosen trapping parameters, the phase transition to a Coulomb crystal is expected to happen at 5~mK, if we use the condition demonstrated for the bulk \cite{pollock73}. As most simulations imply only 100 ions, this transition is expected to happen for a lower temperature. Experimentally, it has been observed that small samples, laser cooled to 4~mK, form an organized and steady structure, even if ions jump from site to site in this structure \cite{haze2012}. It is important here that the kinetic energy of the ions is negligible compared to the Coulomb repulsion when they are ready for transport. The laser cooling is kept during the transport but we estimated that its role is negligible, as the Doppler effect induced by the transport shifts the ions out of resonance.
 
The dynamics simulations show that, once the transport completed, the CM oscillates with an amplitude depending on the transport duration. To have an estimation comparable with the total energy of a single ion, like given by Eq.(\ref{eq:Xi}), the maximum value taken by the CM velocity during the last six oscillations is selected to compute $T_{r,z}^{CM}$. Moreover, the shuttling is expected to induce some discrepancies between the kinetic energy in the radial and axial directions, we have therefore measured both of them separately. The kinetic energy of the CM is converted into a temperature $T_{r,z}^{CM}$, defined by $ k_B T_{r,z}^{CM} = m({v}_{r,z}^{CM})^2 $. 

\begin{figure}
\centering
\includegraphics[width=10cm]{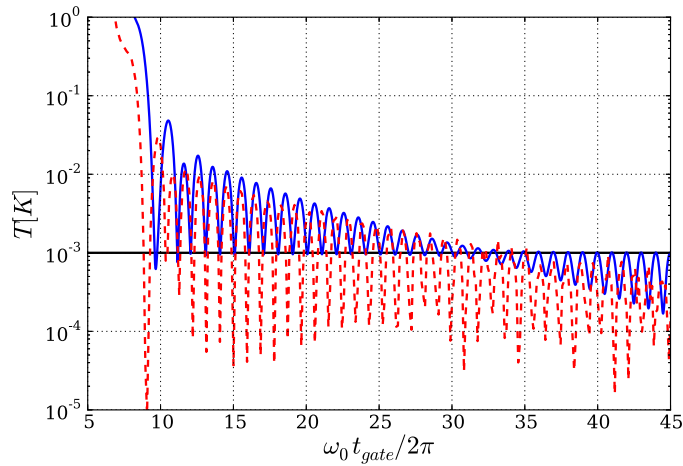}
\caption{(Color online) Energy in the axial direction (converted into a temperature) of a single ion in a 1D translated harmonic potential (blue solid line) and of the CM of a ion cloud ($N_0=100$) in a 3D potential translated along $Oz$ (red dashed line), versus the duration of the transport in reduced units. In both cases, the hyperbolic tangent gate is used. The covered distance is 23~mm and the translated harmonic potential is fixed to $\omega_0/2\pi = 124$~kHz. The effective transverse harmonic potential for the 3D trapping is $\omega_r = 267$~kHz. The single ion starts with an initial velocity equivalent to a temperature of 1~mK and oriented towards the shuttling destination. The ion cloud CM has a negligible initial velocity and the velocity distribution corresponds to a temperature $T(t=0) \approx 4$~mK.}
\label{fig:1D_vs_3D_ideal}
\end{figure}

The energy transferred to the CM of the ion cloud during transport is plotted on figure~\ref{fig:1D_vs_3D_ideal} for comparison with a single ion. We only show the results for a hyperbolic tangent time profile as it is representative of the four considered profiles, except for some aspects that are detailed in the following.

Some conclusions are common to the four time profiles and extrapolate what was demonstrated for two atoms in \cite{palmero13} : for any transport duration, the only observed temperature increase concerns the motion of the CM along the axial direction. The temperature associated to this motion exhibits periodic minima as a function of the transport duration, like for a single ion, for the linear, sinusoidal and hyperbolic tangent profile. In this last case, the first minimum appears for $t_{gate}$ equal to 9 oscillation periods (see figure~\ref{fig:1D_vs_3D_ideal}) whereas it is observable after only one or one and a half period for the two first profiles. For the polynomial profile, the temperature shows no energy gain for all the transport durations explored, like in the case of a single ion with a non null initial velocity. On this figure, the temperature minima for the clouds are lower than the one for the single ion transport because the CM initial velocity is much smaller than the one chosen for the single ion.

These results show that the polynomial gate offers a major advantage as the energy gain cancellation regime does not require specific transport duration and can theoretically be operated for times shorter than one oscillation period. However, if one could produce a moving constant harmonic potential, an ion cloud could be transported with a zero final temperature gain for any of the other time profiles discussed here, provided the appropriate transport duration is chosen.

All the previous analysis assumes a translated harmonic oscillator. In the following section, we present the limitations introduced by the experimental realization of such a transport and we study the effect of such limitations on the ion cloud heating.

\section{Experimental implementation}\label{s:implementation}
The practical realization of an ion-cloud transport is based on the time variation of the voltages applied to the electrodes used for the axial confinement. We will call such a voltage variation a waveform. In order to determine the required waveform, two approaches are found in the literature concerning single ion shuttling. The first one is to search for waveforms which keep the axial trapping potential as close as possible to a harmonic potential with a constant secular frequency throughout the transport. An experimental realization of this approach can be found in \cite{blakestad11}, where a constrained least-squares optimization algorithm is used to transport an ion in a X junction and in \cite{walther12,bowler12} for transport along a linear multizone trap, where the electrodes are sufficiently close to keep the local potential unchanged.

An alternative consists in using optimal control theory as done in \cite{schulz06}. In this approach, the algorithm output is a waveform that minimizes the ion's phase space displacement after the transport. What happens during the transport itself is not relevant (even if constraints can be imposed). Simulations based on the control theory require to solve the equations of motion as many times as required until the algorithm converges. This is computationally feasible for a small number of ions, but for a large ion cloud, the Coulomb interaction is too costly in calculation time to consider such an approach.

These examples \cite{reichle06, schulz06, blakestad11,walther12,bowler12} of shuttling within a well controlled potential were designed in micro-fabricated traps where the typical distance covered by the ion is of the order of few $100 \mu$m thanks to the variation of the voltage applied to a number of electrodes ranging from 4 to 20, with a large size compared to the covered distance. In our experiment, we aim to transport an ion cloud over a distance of 23~mm along an RF quadrupole trap split in two trapping zones by a 2~mm wide extra DC-electrode, setting to 3 the number of available electrodes for transport (see \cite{champenois13} for details on the experimental set-up). Regarding dimensions and external control parameters this is more comparable to the macroscopic set-ups used in mass spectrometry \cite{senko97}, but without the damping role of buffer gas and with the objective of 100~\% shuttling efficiency. In the next subsections, we address the issue of the design of the waveforms, as well as their impact on the ion temperature and ion number after transport.

\subsection{Design of waveforms for ion transport}\label{s:waveform}
\begin{figure}
\centering
\includegraphics[width=10cm]{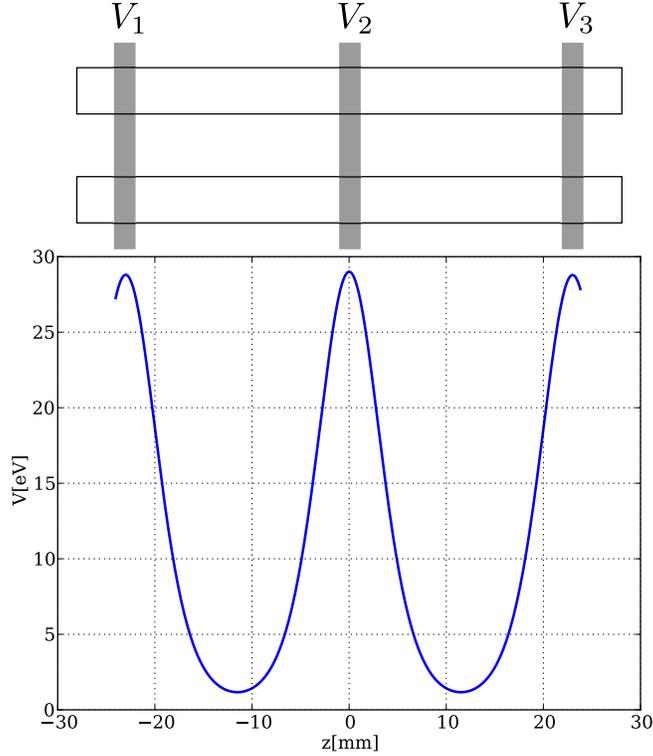}
\caption{ (Color online) Top : Schematic drawing of the double trap. Gray: DC electrodes, white: RF electrodes. Bottom : Axial potential generated by 1000~V applied to the DC electrodes.}
\label{fig:trap1D}
\end{figure}

We now focus on what can be done for transport with three electrodes, the smallest number of electrodes available. Figure \ref{fig:trap1D} shows the trap sketch including the potential distribution and notations. If we call $\phi_i(x,y,z)$ the electric potential created by electrode $i$ when 1V is applied to it, the total DC potential inside the trap can be expressed as \cite{singer10}
\begin{equation}
\Phi(t,x,y,z) = \sum_i^N{V_i(t)\phi_i(x,y,z)}
\end{equation}
if $V_i(t)$ is applied to electrode $i$. In the following, the dependence on $x$ and $y$ is neglected due to the small size of the cloud with respect to the radial size of the trap, therefore $\phi_i(x,y,z)$ can be written as $\phi_{i}(z)$.

 Building a harmonic potential centered on the moving $z_{min}(t)$ implies that
\begin{eqnarray}
\frac{\partial\Phi}{\partial z}\Big|_{z_{min}(t)} &=& 0 \label{eq:cond1}\\
\frac{\partial^{2}\Phi}{\partial z^{2}}\Big|_{z_{min}(t)} &=& \frac{m \omega_0^{2}}{Q} \label{eq:cond2a}
\end{eqnarray}
where $Q$ is the ion charge. Equation~(\ref{eq:cond1}) leads to :
\begin{equation}\label{eq:only_1condition}
V_2(t) = -\frac{V_1(t)\phi^{'}_1(z) + V_3(t)\phi^{'}_3(z)}{\phi^{'}_2(z)} \Big|_{z_{min}(t)}
\end{equation}
where $\phi'_i(z)$ is the first order derivative relative to $z$. Depending on the shape of the electrodes, the $\phi_i(z)$ function can take different forms but they always show a maximum at the center of the electrode which implies that $\phi'_2(z_2)=0$ if we call $z_2$ the center of the second electrode. For any combination of $V_1(t)$ and $V_3(t)$, Eq~(\ref{eq:only_1condition}) leads to a discontinuity of $V_2(t)$ when $z_{min}(t)=z_2$. One can avoid this discontinuity by imposing a constant relation between $V_1(t)$ and $V_3(t)$ given by
\begin{equation}
V_3 (t)= V_1(t) \frac{\phi'_2\phi''_1 - \phi'_1\phi''_2}{\phi'_3\phi''_2 - \phi'_2\phi''_3}\Big|_{z=z_2} \label{eq:V3_1}
\end{equation}
which simplifies to $V_3 (t)= - V_1(t)\phi'_1(z_2)/\phi'_3(z_2)$ if one assumes $\phi'_2(z_2)= 0$, which may not be satisfied in a simulation where the space variables are discretized, as needed in the following. Indeed, the more general expression given by Eq.(\ref{eq:V3_1}) allows the cancellation of the discontinuity even if the space grid does not match exactly the center of the second electrode. Furthermore, in a perfectly symmetric device where the central electrode splits the trap in two identical trapping zones, $\phi'_1(z_2) = -\phi'_3(z_2)$ leading to $V_3(t) = V_1(t)$. Any asymmetry in the electrode environment breaks this equality and the ratio $V_3(t) / V_1(t)$ has to be modified in order to cancel the discontinuity.

In the trap we consider here, the potential $\phi_i(z)$ generated by each electrode has a FWHM of 3.9~mm which is far smaller than the distance between electrodes, equal to 23~mm (cf. Figure \ref{fig:trap1D}). Huge and unrealistic voltages are then required to obey Eq.(\ref{eq:cond2a}). To overcome this impossibility, we study the effect of the deformation of the potential along the transport where the only effective control concerns the position of the potential minimum. In practice, the ratio $V_3 / V_1$ is kept constant and chosen to cancel the discontinuity and $V_2(t)$ is designed such that $z_{min}(t)$ follows the chosen time profile. To have a realistic diagnostic, the molecular dynamics simulation of the cloud is carried out inside the potential generated from the real experimental geometry as given by SIMION8.1 \cite{simion}. The same potential is used to compute $V_3 / V_1$ and $V_2(t)$. To study the effects of the non-stationarity of the potential along the transport, we first expose the single ion case before studying the heating effect on a cloud.

\subsection{Heating of the ions}
\subsubsection{The single ion case}\label{s:1ion}
\begin{figure}
\centering
\includegraphics[width=10cm]{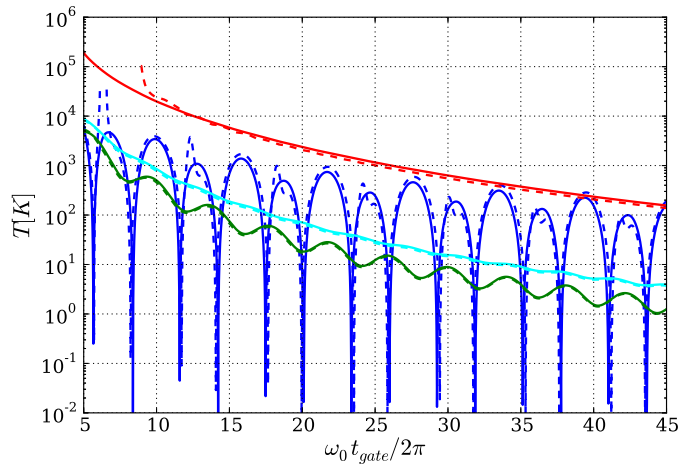}
\caption{(Color online) Energy (converted into temperature) of a single ion shuttled in the real experimental potential (solid lines) with the potential-minimum time profile following a linear gate (blue), a sinus gate (green), a hyperbolic tangent gate (red) and a polynomial gate (cyan), for different transport durations $t_{gate}$ in units of $2\pi/\omega_0$. The initial energy of the ion corresponds to 1~mK (same definition as in Fig.(\ref{fig:1D_vs_3D_ideal})). The dashed lines with same color give the results for a transport in a 1D analytical harmonic potential (Eq.\ref{eq:ideal_potential_w0}) where $\omega_0$ is replaced by $\omega(t)$. }
\label{fig:1D_PA_vs_wt}
\end{figure}

Figure \ref{fig:1D_PA_vs_wt} shows the final energy of an ion transported in a potential computed like explained above, for the four different time profiles defined by Eq.(\ref{eq:lin}-\ref{eq:poly}). We continue to express the transport duration in terms of a harmonic oscillation period $2\pi/\omega_0$ where $\omega_0=\omega_z(t=0)$ and $\omega_z(t)$ is computed thanks to the second derivative of the total axial potential by
\begin{eqnarray}\label{eq:wt}
\omega_z^{2}(t) = \frac{Q}{m} \frac{\partial^{2}\Phi}{\partial z^{2}}\Big|_{z_{min}(t)}
\end{eqnarray}

The ion starts at the potential minimum with an initial velocity equivalent to an energy of 1~mK, oriented towards the transport destination. The periodic minima of the energy gain observed for the translated harmonic potential are now washed out, except for the linear gate, for which a double period is observable. Except for these periodic minima, the final energy is several orders of magnitude higher than the initial one, showing a drastic heating. All gates show a decreasing final energy with longer transport durations, the lowest-lying energy curve is obtained for the sinus gate.

\begin{figure}
\centering
\includegraphics[width=10cm]{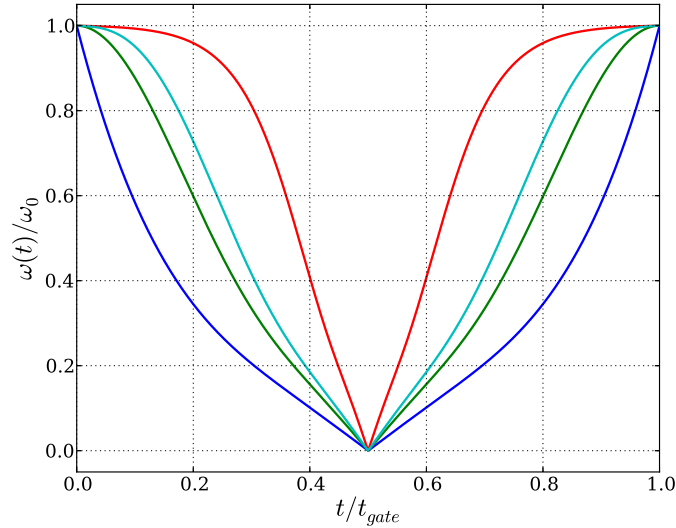}
\caption{(Color online) Relative time evolution of the effective $\omega_z(t)$ versus the relative gate duration $t/t_{gate}$, like calculated with Eq.(\ref{eq:wt}) for the potential-minimum time profile following a linear gate (blue), a sinus gate (green), a hyperbolic tangent gate (red) and a polynomial gate (cyan).}
\label{fig:wt}
\end{figure}

The waveform based transport is characterized by a major difference in the potential evolution compared to the translated harmonic potential transport : the spatial profile of the potential changes along the transport and eventually higher order terms become significant which contribute to the cloud heating. The calculated $\omega_z(t)$ is plotted on figure~\ref{fig:wt} for the four potential minimum time profiles.
Even if the time dependence of $\omega_z(t)$ is different for the four different gates, the molecular dynamics simulation run in an axial potential reduced to its second order contribution show nearly exactly the same features as the one run in the full potential $\Phi(t,x,y,z) $ (see figure \ref{fig:1D_PA_vs_wt}) in the four cases. This comparison demonstrates that the cause of the transport induced heating of a single ion is the time variation of the leading order coefficient, the extra terms having only a very small influence. The preservation of the periodic minima for the linear gate may be explained by its kinetic effect. In practice, a linear gate results in a kick forward at the beginning of the gate and a kick backward at its end resulting in a smaller sensitivity to the intermediate evolution.

The calculated $\omega_z(t)$ shown on figure~\ref{fig:wt} for the four chosen time profiles all reach a value three orders of magnitude smaller than the initial one, at half gate duration, which in practice corresponds to a potential minimum located at the central electrode. This feature, caused by our experimental design, prevents to design an adiabatic transport scheme. Indeed, the adiabaticity criteria for a time varying secular frequency $\omega_z(t)$ without transport is given by $\dot{\omega}/\omega^2 <1$. Due to our trap geometry and trapping parameters, this condition leads to transport durations of 5~s to 60~s, depending on the chosen gate.

\subsubsection{The ion cloud case}
\begin{figure}
\centering
\includegraphics[width=10cm]{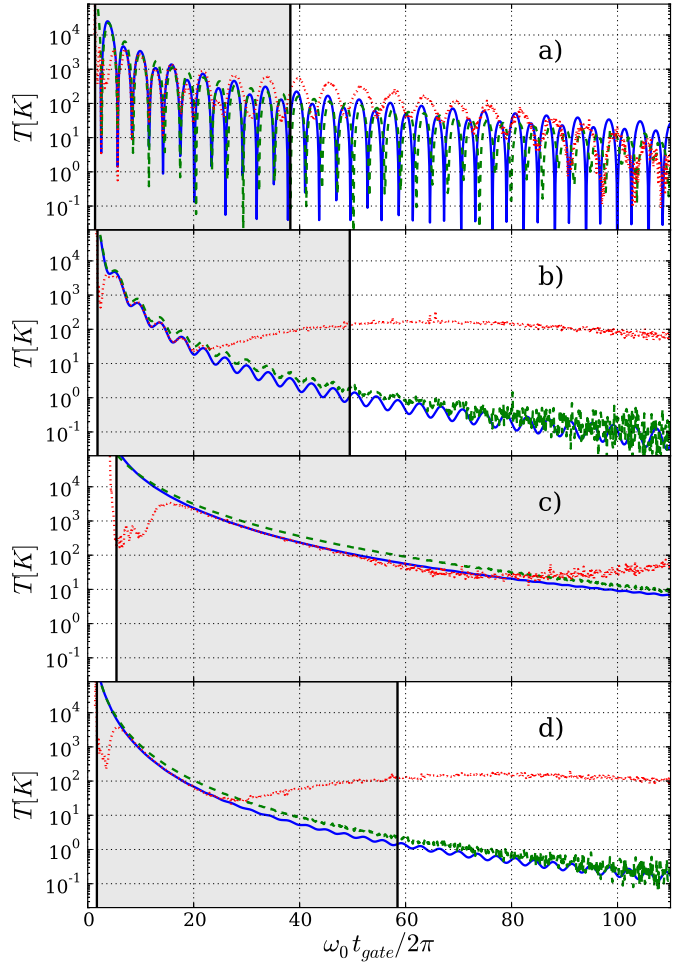}
\caption{(Color online) Axial motion energy of the CM (converted into temperature) of a 100-ion cloud transported in the full 3D moving potential like described in \ref{s:waveform} (red dotted line), versus the gate duration in reduced unit $\omega_0t_{gate}/2\pi $. For comparison, the single ion case, as described in Figure \ref{fig:1D_PA_vs_wt}, is indicated by the solid blue and the dashed green lines show the results for the same cloud when only the harmonic contribution of the axial potential is kept, (see Eq.(\ref{eq:wt})). The different minimum time profiles obey to a linear gate (a), a sinus gate (b), a hyperbolic tangent gate (c) and a polynomial gate (d). The shaded areas indicate the durations for which 100\% of the ions are transported. }
\label{fig:transport_1mK_PA}
\end{figure}

\begin{figure}
\centering
\includegraphics[width=10cm]{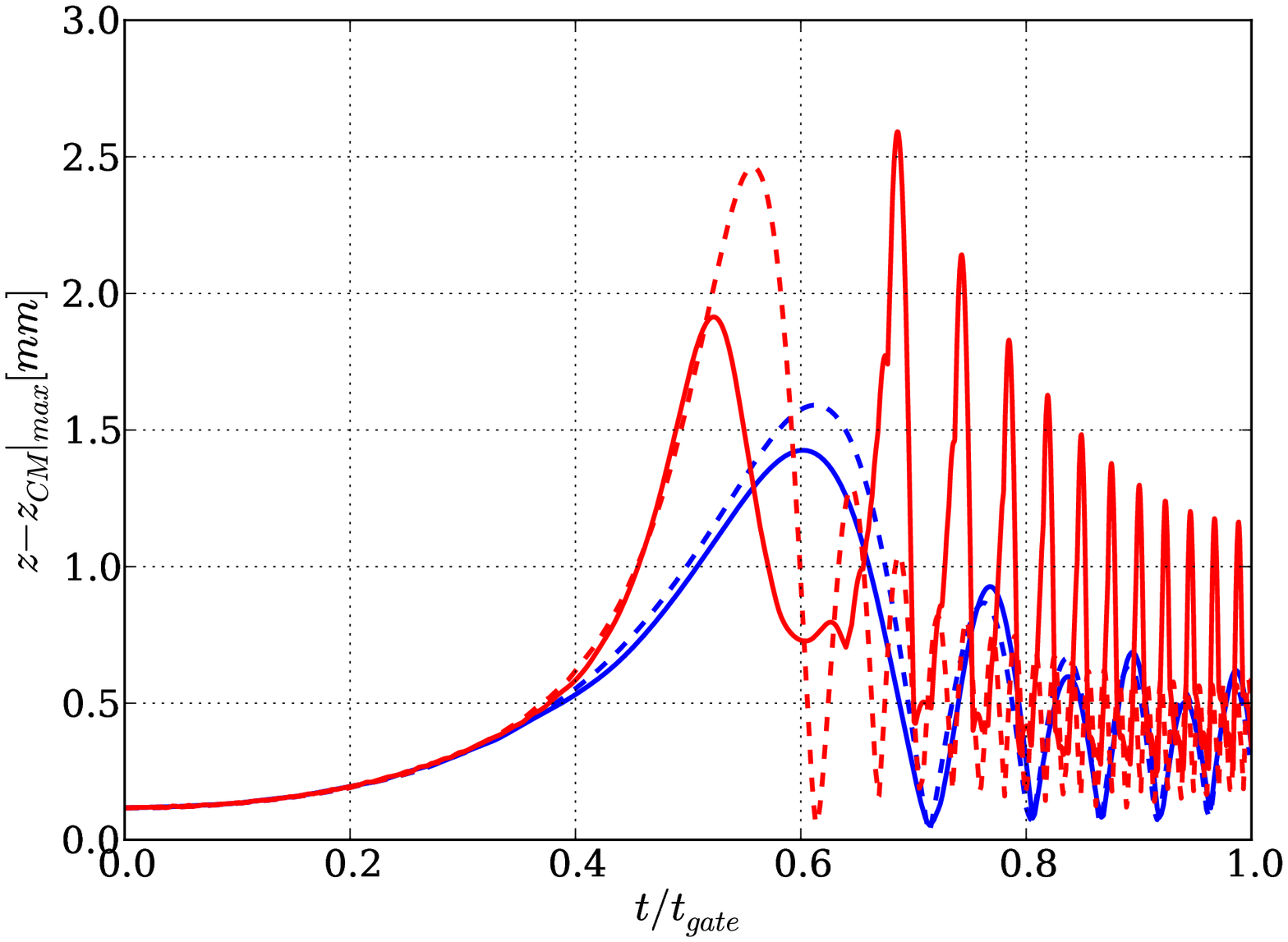}
\caption{(Color online) Spatial extension (largest value of $z_i-z_{CM}$) of a 100-ion cloud transported in the full 3D moving potential (solid lines) and in its harmonic contribution (dashed lines), versus relative time ($t / t_{gate}$), for a gate of duration 100~$\mu$s (blue lines) and 400~$\mu$s (red lines). The potential-minimum time profile obeys the sinus gate.}
\label{fig:extension}
\end{figure}

Considering the case of the ion cloud, we compare the final energy of the axial motion of the cloud's CM to the energy of a single ion transported in the same potential. Figure \ref{fig:transport_1mK_PA} shows that their behavior with the transport duration is essentially the same, provided that only the harmonic contribution to the axial potential is kept. This property is shared by the four tested profile gates and confirms a net increase of the CM energy induced by the transport, like already observed for the single ion case. When the simulations are run in the potential given by SIMION, figure \ref{fig:transport_1mK_PA} shows that for long enough gates (except for the linear one) the evolution of the energy in the cloud case differs from the single ion case. Indeed, after an initial decrease, the final CM energy curve splits from the single-ion case, deviating to higher values.
The comparison of the sinus, tanh and polynomial gates shows that the two curves split when the energy in the CM motion reaches a value of the order of 40~K, which happens for very different transport durations depending on the chosen gate. This increase of the final energy of the CM makes long transport inappropriate for ion clouds, except with a linear gate (which presents other drawbacks presented in section \ref{s:efficiency}). The higher order terms in the full axial potential probably play a role for long transport durations because of the spatial spreading of the cloud, larger for longer transport. Evidence of this can be found in Figure~\ref{fig:extension}, where the spreading of the cloud is analyzed by plotting the maximum distance between the CM and an ion in the cloud. For this figure, we choose two characteristic durations for transport by a sinus gate : one for which the numerical simulations give approximately the same final temperature for both potentials ($t_{gate}=100~\mu$s, $\omega_0 t_{gate}/2\pi=12.4$) and one which results in different final energy ($t_{gate}=400~\mu$s, $\omega_0 t_{gate}/2\pi=49.6$).

First, we observe that the spreading of the cloud along a transport during 100~$\mu$s is very similar in the full potential and in its harmonic contribution. The oscillations are the signature of the cloud breathing around its CM. The first breathing reaches maximum amplitude which happens around 60\% of the transport duration. The situation is very different for the 400~$\mu$s transport. In this case, the breathing starts relatively sooner during the transport and the breathing in the full potential is rapidly out of phase with the one observed in the harmonic potential. The perturbations induced by the extra terms of the potential amplify the oscillation amplitude which translates into kinetic energy in the CM frame, which we refer to as internal energy. We assume that the non-harmonic terms are responsible for a coupling between internal and CM degrees of motion because the kinetic energy increase is relatively more pronounced on the CM motion than on the internal energy.

Concerning the evolution of the internal energy, it can be resumed as follows: the axial internal kinetic energy fluctuates around a mean value close to 100~K for long enough transport durations, independently of the transport duration and the gate used and for both cases, the full potential and its harmonic contribution. 

Let's notice that in case of a transport in the full potential, this internal kinetic energy reaches very high values, as large as 10000~K , for short transport duration. It takes durations ranging from 50~$\mu$s for the sinus and polynomial gates, to 200~$\mu$s for the linear and tanh gates to reach the 100~K value.

Concerning the radial internal kinetic energy, no drastic heating is observed for fast transports and the calculations give a mean value of 10~K for all gates in the full potential simulations, it can reach 30~K if only the harmonic contribution is used. If we analyze the same internal kinetic energy in a translated constant harmonic potential, we observe no increase at all of the internal kinetic energy which remains equal to 4~mK, like set by Doppler cooling (see \ref{s:3D_w0}).

Furthermore, if we keep the potential minimum immobile but apply to the trap a time variation $\omega_z(t)$ like given by Eq.(\ref{eq:wt}) identical to what is encountered in the moving full potential (see Fig~\ref{fig:wt}), we observe exactly the same evolution of the internal energy, all along the duration of the gate. We can conclude that it is the deformation of the harmonic part of the axial potential that is responsible for the internal heating of the cloud along its transport.

The radial and axial motion of the CM do not seem to couple as its radial energy ranges between 1 and 10~mK in the full potential after transport. The largest values (10~mK) are observed for the linear gate for any transport duration and for transport faster than 50~$\mu$s for the sinus and polynomial gates, and faster than 200~$\mu$s for the tanh gates (100~mK). The same characteristic transport durations were already identified for the internal axial kinetic energy. This small amount of energy is transferred to the radial motion through the non-harmonic part of the potential as it is as low as 0.1~mK if the transport is computed in the harmonic part of the potential, for any gate of any duration.

\begin{figure}
\centering
\includegraphics[width=10cm]{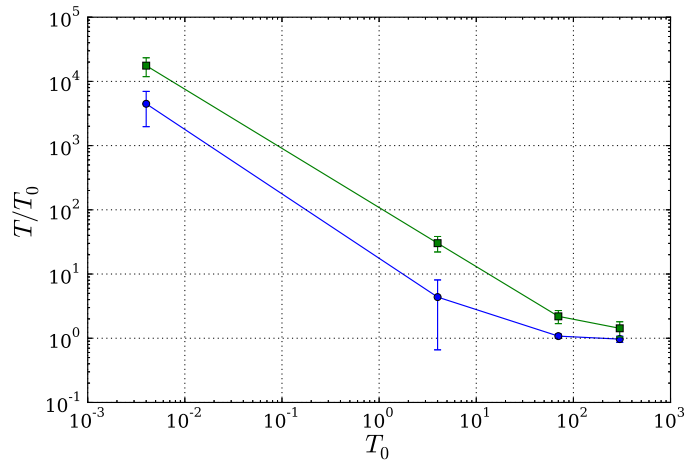}
\caption{(Color online) Variation of the internal temperature versus initial temperature for axial (blue squares) and radial (green circles) motion.}
\label{fig:effect_T0}
\end{figure}

Figure~\ref{fig:effect_T0} shows the transport-induced variation of the internal temperature  as a function of the initial temperature $T_{0}$ for 4 characteristic values : 4~mK, 4~K, 77~K and 300~K. We estimate that this variation can be  explained by a transfer of Coulomb energy to kinetic energy, and therefore does not play an important role for temperatures above some tens of Kelvin.

In this section, we have addressed the heating issue, to enlighten the analogies and discrepancies with a single ion shuttling. In the next section, we focus on the other major concern with ion cloud transport : the transport efficiency.

\subsection{Transport efficiencies}\label{s:efficiency}
The drastic heating mentioned for very short transport duration does not necessarily induce an ion loss in the potential we simulate, whose depth is $\approx$28~eV~$= 336 \cdot 10^{3} K$ in the axial direction and 9.0~eV $= 1.1 \cdot 10^{6} K$ in the radial direction. The transport efficiency is defined as the number of ions trapped in the destination part of the double trap set-up, once the transport is completed. The limits of 100 \% efficient transport are shown by the shaded area in fig~\ref{fig:transport_1mK_PA}. Its lower boundary is due to the inertia of the ions. If the transport is very fast, analysis of the cloud dynamics shows that the ions never leave the first trap, as the gating potential ($V_2$) varies too fast for ions to move to the other part of the trap before the central potential is raised again. This behaviour is also observed for a single ion and thus does not depend on the number of trapped ions. The upper boundary for 100\% transport is due to the fact that for too long gates the ion cloud spreads so much in the axial direction that some ions do not leave the initial part of the trap. The comparison between the different gates of figure~\ref{fig:transport_1mK_PA} shows that this limit depends on the minimum time profile and is out of the scale of this figure for the hyperbolic tangent gate and for 100 ions. 

\begin{figure}
\centering
\includegraphics[width =10cm]{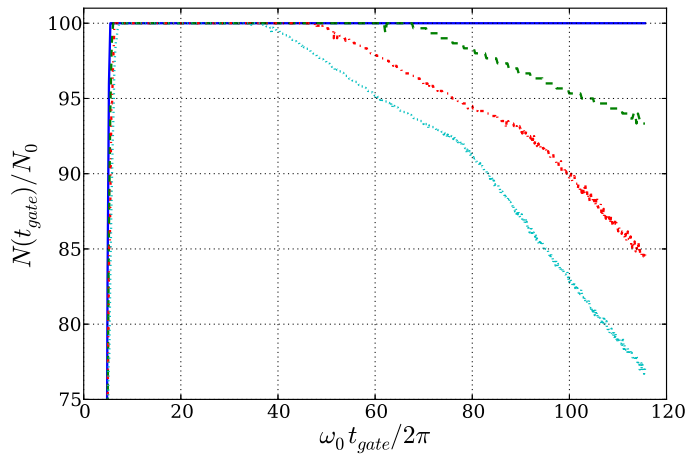}
\caption{(Color online) Transport efficiency versus the transport duration $t_{gate}$ in units of $2\pi/\omega_0$, for an hyperbolic tangent gate and different initial ion number. The initial internal temperature is $\approx 4$~mK in each case. Blue solid line:$N_{0}=100$. Green dashed line: $N_{0}=300$. Red dash-dot line $N_{0}=600$, Cyan dotted line $N_{0}=1000$}
\label{fig:N_efficiency_for_diff_N0}
\end{figure}

\begin{figure}
\centering
\includegraphics[width = 10cm]{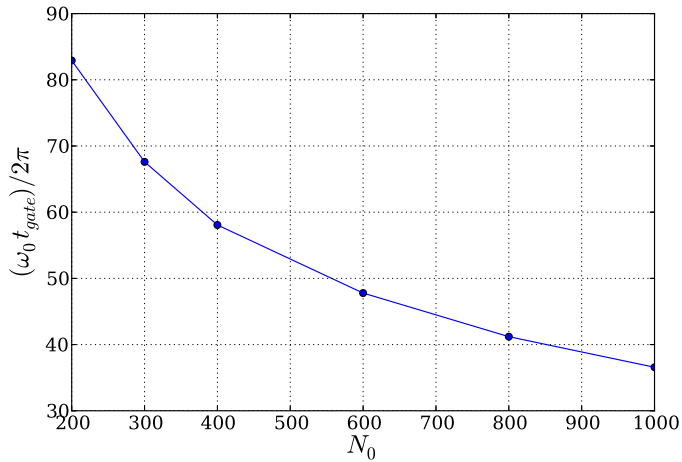}
\caption{(Color online) Transport duration $t_{gate}$ at which the transport efficiency is no longer 100\% for an hyperbolic tangent gate and different initial ion number.}
\label{fig:tg_efficiency}
\end{figure}

A comparison in transport efficiency for different initial ion numbers is made for the hyperbolic tangent gate in Figure~\ref{fig:N_efficiency_for_diff_N0}. After some gate duration, the efficiency diminish linearly with $t_{gate}$, with an apparent change of slope. For an increasing number of ions, the long transport efficiency decreases more rapidly. This behaviour is characteristic of a number dependent non-linear effect and might be due to space charge effects \cite{alheit95}. If we plot the transport time at which the efficiency is no longer 100\% for different $N_0$ (see figure~\ref{fig:tg_efficiency}), we observe a duration reduction indicating that for large ion clouds, there may be no transport duration for which 100\% of the ions would be shuttled. 

\section{Conclusion}
In conclusion, we have carried out numerical simulations to identify the energy gain source and the cause of reduced efficiency in the shuttling of an ion cloud between the two parts of a linear RF traps separated by 2.3~cm. To compare our results with single ion experiments, we limited this study to four potential minimum time profiles already studied in the frame of single ion shuttling. Our transport protocol is limited by the distance between the electrodes controlling the translated trapping potential which induce deformation of this potential along the transport. Two causes of transport energy gain are identified. First, the deformation of the harmonic part of the potential is responsible for a large gain in the axial motion kinetic energy of the CM and for the washing out of the periodic energy gain cancellation, except for a linear potential minimum time profile. This deformation is also responsible for the gain in the kinetic energy in the CM frame for the four studied time profiles. Increasing the transport duration does not lead to a reduction of the CM energy gain because of the non-harmonic part of the potential which plays a role for long transport.

Nevertheless, an efficient (100\%) and fast ($200\mu s$) transport of an ion cloud made of more than 1000 ions is possible at the expense of an energy gain very detrimental for cold ions but irrelevant for clouds with temperatures above 4~K. A challenge is now to design a gate that takes into account the time variation of the secular frequency to reduce the transport induced energy gain.

\begin{acknowledgments}
Scientific discussions with L. Hilico are gratefully acknowledged. This work was partially carried out under ANR grant ANR-08-0053-01. We also acknowledge financial support from CNES and from Region PACA (DiMGAF-2008/22485)

\end{acknowledgments}


\begin{thebibliography}{32}%
\makeatletter
\providecommand \@ifxundefined [1]{%
 \@ifx{#1\undefined}
}%
\providecommand \@ifnum [1]{%
 \ifnum #1\expandafter \@firstoftwo
 \else \expandafter \@secondoftwo
 \fi
}%
\providecommand \@ifx [1]{%
 \ifx #1\expandafter \@firstoftwo
 \else \expandafter \@secondoftwo
 \fi
}%
\providecommand \natexlab [1]{#1}%
\providecommand \enquote  [1]{``#1''}%
\providecommand \bibnamefont  [1]{#1}%
\providecommand \bibfnamefont [1]{#1}%
\providecommand \citenamefont [1]{#1}%
\providecommand \href@noop [0]{\@secondoftwo}%
\providecommand \href [0]{\begingroup \@sanitize@url \@href}%
\providecommand \@href[1]{\@@startlink{#1}\@@href}%
\providecommand \@@href[1]{\endgroup#1\@@endlink}%
\providecommand \@sanitize@url [0]{\catcode `\\12\catcode `\$12\catcode
  `\&12\catcode `\#12\catcode `\^12\catcode `\_12\catcode `\%12\relax}%
\providecommand \@@startlink[1]{}%
\providecommand \@@endlink[0]{}%
\providecommand \url  [0]{\begingroup\@sanitize@url \@url }%
\providecommand \@url [1]{\endgroup\@href {#1}{\urlprefix }}%
\providecommand \urlprefix  [0]{URL }%
\providecommand \Eprint [0]{\href }%
\providecommand \doibase [0]{http://dx.doi.org/}%
\providecommand \selectlanguage [0]{\@gobble}%
\providecommand \bibinfo  [0]{\@secondoftwo}%
\providecommand \bibfield  [0]{\@secondoftwo}%
\providecommand \translation [1]{[#1]}%
\providecommand \BibitemOpen [0]{}%
\providecommand \bibitemStop [0]{}%
\providecommand \bibitemNoStop [0]{.\EOS\space}%
\providecommand \EOS [0]{\spacefactor3000\relax}%
\providecommand \BibitemShut  [1]{\csname bibitem#1\endcsname}%
\let\auto@bib@innerbib\@empty
\bibitem [{\citenamefont {Teloy}\ and\ \citenamefont
  {Gerlich}(1974)}]{teloy74}%
  \BibitemOpen
  \bibfield  {author} {\bibinfo {author} {\bibfnamefont {E.}~\bibnamefont
  {Teloy}}\ and\ \bibinfo {author} {\bibfnamefont {D.}~\bibnamefont
  {Gerlich}},\ }\href {\doibase 10.1016/0301-0104(74)85008-1} {\bibfield
  {journal} {\bibinfo  {journal} {Chemical Physics}\ }\textbf {\bibinfo
  {volume} {4}},\ \bibinfo {pages} {417 } (\bibinfo {year} {1974})}\BibitemShut
  {NoStop}%
\bibitem [{\citenamefont {Senko}\ \emph {et~al.}(1997)\citenamefont {Senko},
  \citenamefont {Hendrickson}, \citenamefont {Emmett}, \citenamefont {Shi},\
  and\ \citenamefont {Marshall}}]{senko97}%
  \BibitemOpen
  \bibfield  {author} {\bibinfo {author} {\bibfnamefont {M.~W.}\ \bibnamefont
  {Senko}}, \bibinfo {author} {\bibfnamefont {C.~L.}\ \bibnamefont
  {Hendrickson}}, \bibinfo {author} {\bibfnamefont {M.~R.}\ \bibnamefont
  {Emmett}}, \bibinfo {author} {\bibfnamefont {S.~D.-H.}\ \bibnamefont {Shi}},
  \ and\ \bibinfo {author} {\bibfnamefont {A.~G.}\ \bibnamefont {Marshall}},\
  }\href {\doibase http://dx.doi.org/10.1016/S1044-0305(97)00126-8} {\bibfield
  {journal} {\bibinfo  {journal} {Journal of the American Society for Mass
  Spectrometry}\ }\textbf {\bibinfo {volume} {8}},\ \bibinfo {pages} {970 }
  (\bibinfo {year} {1997})}\BibitemShut {NoStop}%
\bibitem [{\citenamefont {Herfurth}(2003)}]{herfurth03}%
  \BibitemOpen
  \bibfield  {author} {\bibinfo {author} {\bibfnamefont {F.}~\bibnamefont
  {Herfurth}},\ }\href {\doibase 10.1016/S0168-583X(02)02135-3} {\bibfield
  {journal} {\bibinfo  {journal} {Nuclear Instruments and Methods in Physics
  Research Section B: Beam Interactions with Materials and Atoms}\ }\textbf
  {\bibinfo {volume} {204}},\ \bibinfo {pages} {587 } (\bibinfo {year}
  {2003})},\ \bibinfo {note} {14th International Conference on Electromagnetic
  Isotope Separators and Techniques Related to their Applications}\BibitemShut
  {NoStop}%
\bibitem [{\citenamefont {Haegg}\ and\ \citenamefont {Szabo}(1986)}]{haegg86}%
  \BibitemOpen
  \bibfield  {author} {\bibinfo {author} {\bibfnamefont {C.}~\bibnamefont
  {Haegg}}\ and\ \bibinfo {author} {\bibfnamefont {I.}~\bibnamefont {Szabo}},\
  }\href {\doibase 10.1016/0168-1176(86)80004-0} {\bibfield  {journal}
  {\bibinfo  {journal} {International Journal of Mass Spectrometry and Ion
  Processes}\ }\textbf {\bibinfo {volume} {73}},\ \bibinfo {pages} {295 }
  (\bibinfo {year} {1986})}\BibitemShut {NoStop}%
\bibitem [{\citenamefont {Kielpinski}\ \emph {et~al.}(2002)\citenamefont
  {Kielpinski}, \citenamefont {Monroe},\ and\ \citenamefont
  {Wineland}}]{kielpinski02}%
  \BibitemOpen
  \bibfield  {author} {\bibinfo {author} {\bibfnamefont {D.}~\bibnamefont
  {Kielpinski}}, \bibinfo {author} {\bibfnamefont {C.}~\bibnamefont {Monroe}},
  \ and\ \bibinfo {author} {\bibfnamefont {D.}~\bibnamefont {Wineland}},\
  }\href@noop {} {\bibfield  {journal} {\bibinfo  {journal} {Nature}\ }\textbf
  {\bibinfo {volume} {417}},\ \bibinfo {pages} {709} (\bibinfo {year}
  {2002})}\BibitemShut {NoStop}%
\bibitem [{\citenamefont {Walther}\ \emph {et~al.}(2012)\citenamefont
  {Walther}, \citenamefont {Ziesel}, \citenamefont {Ruster}, \citenamefont
  {Dawkins}, \citenamefont {Ott}, \citenamefont {Hettrich}, \citenamefont
  {Singer}, \citenamefont {Schmidt-Kaler},\ and\ \citenamefont
  {Poschinger}}]{walther12}%
  \BibitemOpen
  \bibfield  {author} {\bibinfo {author} {\bibfnamefont {A.}~\bibnamefont
  {Walther}}, \bibinfo {author} {\bibfnamefont {F.}~\bibnamefont {Ziesel}},
  \bibinfo {author} {\bibfnamefont {T.}~\bibnamefont {Ruster}}, \bibinfo
  {author} {\bibfnamefont {S.~T.}\ \bibnamefont {Dawkins}}, \bibinfo {author}
  {\bibfnamefont {K.}~\bibnamefont {Ott}}, \bibinfo {author} {\bibfnamefont
  {M.}~\bibnamefont {Hettrich}}, \bibinfo {author} {\bibfnamefont
  {K.}~\bibnamefont {Singer}}, \bibinfo {author} {\bibfnamefont
  {F.}~\bibnamefont {Schmidt-Kaler}}, \ and\ \bibinfo {author} {\bibfnamefont
  {U.}~\bibnamefont {Poschinger}},\ }\href {\doibase
  10.1103/PhysRevLett.109.080501} {\bibfield  {journal} {\bibinfo  {journal}
  {Phys. Rev. Lett.}\ }\textbf {\bibinfo {volume} {109}},\ \bibinfo {pages}
  {080501} (\bibinfo {year} {2012})}\BibitemShut {NoStop}%
\bibitem [{\citenamefont {Bowler}\ \emph {et~al.}(2012)\citenamefont {Bowler},
  \citenamefont {Gaebler}, \citenamefont {Lin}, \citenamefont {Tan},
  \citenamefont {Hanneke}, \citenamefont {Jost}, \citenamefont {Home},
  \citenamefont {Leibfried},\ and\ \citenamefont {Wineland}}]{bowler12}%
  \BibitemOpen
  \bibfield  {author} {\bibinfo {author} {\bibfnamefont {R.}~\bibnamefont
  {Bowler}}, \bibinfo {author} {\bibfnamefont {J.}~\bibnamefont {Gaebler}},
  \bibinfo {author} {\bibfnamefont {Y.}~\bibnamefont {Lin}}, \bibinfo {author}
  {\bibfnamefont {T.~R.}\ \bibnamefont {Tan}}, \bibinfo {author} {\bibfnamefont
  {D.}~\bibnamefont {Hanneke}}, \bibinfo {author} {\bibfnamefont {J.~D.}\
  \bibnamefont {Jost}}, \bibinfo {author} {\bibfnamefont {J.~P.}\ \bibnamefont
  {Home}}, \bibinfo {author} {\bibfnamefont {D.}~\bibnamefont {Leibfried}}, \
  and\ \bibinfo {author} {\bibfnamefont {D.~J.}\ \bibnamefont {Wineland}},\
  }\href {\doibase 10.1103/PhysRevLett.109.080502} {\bibfield  {journal}
  {\bibinfo  {journal} {Phys. Rev. Lett.}\ }\textbf {\bibinfo {volume} {109}},\
  \bibinfo {pages} {080502} (\bibinfo {year} {2012})}\BibitemShut {NoStop}%
\bibitem [{\citenamefont {Couvert}\ \emph {et~al.}(2008)\citenamefont
  {Couvert}, \citenamefont {Kawalec}, \citenamefont {Reinaudi},\ and\
  \citenamefont {Guéry-Odelin}}]{couvert08}%
  \BibitemOpen
  \bibfield  {author} {\bibinfo {author} {\bibfnamefont {A.}~\bibnamefont
  {Couvert}}, \bibinfo {author} {\bibfnamefont {T.}~\bibnamefont {Kawalec}},
  \bibinfo {author} {\bibfnamefont {G.}~\bibnamefont {Reinaudi}}, \ and\
  \bibinfo {author} {\bibfnamefont {D.}~\bibnamefont {Guéry-Odelin}},\ }\href
  {\doibase 10.1209/0295-5075/83/13001} {\bibfield  {journal} {\bibinfo
  {journal} {{EPL} (Europhysics Letters)}\ }\textbf {\bibinfo {volume} {83}},\
  \bibinfo {pages} {13001} (\bibinfo {year} {2008})}\BibitemShut {NoStop}%
\bibitem [{\citenamefont {Schaff}\ \emph {et~al.}(2011)\citenamefont {Schaff},
  \citenamefont {Song}, \citenamefont {Capuzzi}, \citenamefont {Vignolo},\ and\
  \citenamefont {Labeyrie}}]{schaff11a}%
  \BibitemOpen
  \bibfield  {author} {\bibinfo {author} {\bibfnamefont {J.-F.}\ \bibnamefont
  {Schaff}}, \bibinfo {author} {\bibfnamefont {X.-L.}\ \bibnamefont {Song}},
  \bibinfo {author} {\bibfnamefont {P.}~\bibnamefont {Capuzzi}}, \bibinfo
  {author} {\bibfnamefont {P.}~\bibnamefont {Vignolo}}, \ and\ \bibinfo
  {author} {\bibfnamefont {G.}~\bibnamefont {Labeyrie}},\ }\href
  {http://stacks.iop.org/0295-5075/93/i=2/a=23001} {\bibfield  {journal}
  {\bibinfo  {journal} {EPL (Europhysics Letters)}\ }\textbf {\bibinfo {volume}
  {93}},\ \bibinfo {pages} {23001} (\bibinfo {year} {2011})}\BibitemShut
  {NoStop}%
\bibitem [{\citenamefont {Pedregosa}\ \emph {et~al.}(2010)\citenamefont
  {Pedregosa}, \citenamefont {Champenois}, \citenamefont {Houssin},\ and\
  \citenamefont {Knoop}}]{pedregosa10a}%
  \BibitemOpen
  \bibfield  {author} {\bibinfo {author} {\bibfnamefont {J.}~\bibnamefont
  {Pedregosa}}, \bibinfo {author} {\bibfnamefont {C.}~\bibnamefont
  {Champenois}}, \bibinfo {author} {\bibfnamefont {M.}~\bibnamefont {Houssin}},
  \ and\ \bibinfo {author} {\bibfnamefont {M.}~\bibnamefont {Knoop}},\ }\href
  {\doibase DOI: 10.1016/j.ijms.2009.12.009} {\bibfield  {journal} {\bibinfo
  {journal} {International Journal of Mass Spectrometry}\ }\textbf {\bibinfo
  {volume} {290}},\ \bibinfo {pages} {100 } (\bibinfo {year}
  {2010})}\BibitemShut {NoStop}%
\bibitem [{\citenamefont {Champenois}\ \emph {et~al.}(2013)\citenamefont
  {Champenois}, \citenamefont {Pedregosa-Gutierrez}, \citenamefont {Marciante},
  \citenamefont {Guyomarc'h}, \citenamefont {Houssin},\ and\ \citenamefont
  {Knoop}}]{champenois13}%
  \BibitemOpen
  \bibfield  {author} {\bibinfo {author} {\bibfnamefont {C.}~\bibnamefont
  {Champenois}}, \bibinfo {author} {\bibfnamefont {J.}~\bibnamefont
  {Pedregosa-Gutierrez}}, \bibinfo {author} {\bibfnamefont {M.}~\bibnamefont
  {Marciante}}, \bibinfo {author} {\bibfnamefont {D.}~\bibnamefont
  {Guyomarc'h}}, \bibinfo {author} {\bibfnamefont {M.}~\bibnamefont {Houssin}},
  \ and\ \bibinfo {author} {\bibfnamefont {M.}~\bibnamefont {Knoop}},\ }\href
  {\doibase http://dx.doi.org/10.1063/1.4796077} {\bibfield  {journal}
  {\bibinfo  {journal} {AIP Conference Proceedings}\ }\textbf {\bibinfo
  {volume} {1521}},\ \bibinfo {pages} {210} (\bibinfo {year}
  {2013})}\BibitemShut {NoStop}%
\bibitem [{\citenamefont {Prestage}\ \emph {et~al.}(2003)\citenamefont
  {Prestage}, \citenamefont {Chung}, \citenamefont {Le}, \citenamefont {Beach},
  \citenamefont {Maleki},\ and\ \citenamefont {Tjoelker}}]{prestage03}%
  \BibitemOpen
  \bibfield  {author} {\bibinfo {author} {\bibfnamefont {J.}~\bibnamefont
  {Prestage}}, \bibinfo {author} {\bibfnamefont {S.}~\bibnamefont {Chung}},
  \bibinfo {author} {\bibfnamefont {T.}~\bibnamefont {Le}}, \bibinfo {author}
  {\bibfnamefont {M.}~\bibnamefont {Beach}}, \bibinfo {author} {\bibfnamefont
  {L.}~\bibnamefont {Maleki}}, \ and\ \bibinfo {author} {\bibfnamefont
  {R.}~\bibnamefont {Tjoelker}},\ }\href@noop {} {\bibfield  {journal}
  {\bibinfo  {journal} {Proceedings of 35$^{th}$ Annual Precise Time and Time
  Interval (PTTI) Meeting}\ } (\bibinfo {year} {2003})},\ \bibinfo {note}
  {http://tycho.usno.navy.mil/ptti/ptti2003/paper40.pdf}\BibitemShut {NoStop}%
\bibitem [{\citenamefont {Prestage}(1995)}]{prestage95}%
  \BibitemOpen
  \bibfield  {author} {\bibinfo {author} {\bibfnamefont {J.}~\bibnamefont
  {Prestage}},\ }\href
  {http://www.google.com/patents?hl=en\&lr=\&vid=USPAT5420549\&id=rLgdAAAAEBAJ\&oi=fnd\&dq=%22J+prestage%22\&printsec=abstract}
  {\bibfield  {journal} {\bibinfo  {journal} {US Patent 5,420,549}\ } (\bibinfo
  {year} {1995})}\BibitemShut {NoStop}%
\bibitem [{\citenamefont {Dicke}(1953)}]{dicke53}%
  \BibitemOpen
  \bibfield  {author} {\bibinfo {author} {\bibfnamefont {R.~H.}\ \bibnamefont
  {Dicke}},\ }\href@noop {} {\bibfield  {journal} {\bibinfo  {journal} {Phys.
  Rev.}\ }\textbf {\bibinfo {volume} {89}},\ \bibinfo {pages} {472} (\bibinfo
  {year} {1953})}\BibitemShut {NoStop}%
\bibitem [{\citenamefont {et. al.}(2014)}]{kamsap14}%
  \BibitemOpen
  \bibfield  {author} {\bibinfo {author} {\bibfnamefont {M.~R.~Kamsap}\
  \bibnamefont {et. al.}},\ }\href@noop {} {} (\bibinfo {year} {2014}),\
  \bibinfo {note} {in preparation}\BibitemShut {NoStop}%
\bibitem [{\citenamefont {Bl\"{u}mel}\ \emph {et~al.}(1988)\citenamefont
  {Bl\"{u}mel}, \citenamefont {Chen}, \citenamefont {Peik}, \citenamefont
  {Quint}, \citenamefont {Schleich}, \citenamefont {Shen},\ and\ \citenamefont
  {Walther}}]{bluemel88}%
  \BibitemOpen
  \bibfield  {author} {\bibinfo {author} {\bibfnamefont {R.}~\bibnamefont
  {Bl\"{u}mel}}, \bibinfo {author} {\bibfnamefont {J.~M.}\ \bibnamefont
  {Chen}}, \bibinfo {author} {\bibfnamefont {E.}~\bibnamefont {Peik}}, \bibinfo
  {author} {\bibfnamefont {W.}~\bibnamefont {Quint}}, \bibinfo {author}
  {\bibfnamefont {W.}~\bibnamefont {Schleich}}, \bibinfo {author}
  {\bibfnamefont {Y.~R.}\ \bibnamefont {Shen}}, \ and\ \bibinfo {author}
  {\bibfnamefont {H.}~\bibnamefont {Walther}},\ }\href@noop {} {\bibfield
  {journal} {\bibinfo  {journal} {Nature}\ }\textbf {\bibinfo {volume} {334}},\
  \bibinfo {pages} {309 } (\bibinfo {year} {1988})}\BibitemShut {NoStop}%
\bibitem [{\citenamefont {Marciante}\ \emph {et~al.}(2010)\citenamefont
  {Marciante}, \citenamefont {Champenois}, \citenamefont {Calisti},
  \citenamefont {Pedregosa-Gutierrez},\ and\ \citenamefont
  {Knoop}}]{marciante10}%
  \BibitemOpen
  \bibfield  {author} {\bibinfo {author} {\bibfnamefont {M.}~\bibnamefont
  {Marciante}}, \bibinfo {author} {\bibfnamefont {C.}~\bibnamefont
  {Champenois}}, \bibinfo {author} {\bibfnamefont {A.}~\bibnamefont {Calisti}},
  \bibinfo {author} {\bibfnamefont {J.}~\bibnamefont {Pedregosa-Gutierrez}}, \
  and\ \bibinfo {author} {\bibfnamefont {M.}~\bibnamefont {Knoop}},\ }\href
  {\doibase 10.1103/PhysRevA.82.033406} {\bibfield  {journal} {\bibinfo
  {journal} {Phys. Rev. A}\ }\textbf {\bibinfo {volume} {82}},\ \bibinfo
  {pages} {033406} (\bibinfo {year} {2010})}\BibitemShut {NoStop}%
\bibitem [{\citenamefont {Palmero}\ \emph {et~al.}(2013)\citenamefont
  {Palmero}, \citenamefont {Torrontegui}, \citenamefont {Gu\'ery-Odelin},\ and\
  \citenamefont {Muga}}]{palmero13}%
  \BibitemOpen
  \bibfield  {author} {\bibinfo {author} {\bibfnamefont {M.}~\bibnamefont
  {Palmero}}, \bibinfo {author} {\bibfnamefont {E.}~\bibnamefont
  {Torrontegui}}, \bibinfo {author} {\bibfnamefont {D.}~\bibnamefont
  {Gu\'ery-Odelin}}, \ and\ \bibinfo {author} {\bibfnamefont {J.~G.}\
  \bibnamefont {Muga}},\ }\href {\doibase 10.1103/PhysRevA.88.053423}
  {\bibfield  {journal} {\bibinfo  {journal} {Phys. Rev. A}\ }\textbf {\bibinfo
  {volume} {88}},\ \bibinfo {pages} {053423} (\bibinfo {year}
  {2013})}\BibitemShut {NoStop}%
\bibitem [{\citenamefont {Reichle}\ \emph {et~al.}(2006)\citenamefont
  {Reichle}, \citenamefont {Leibfried}, \citenamefont {Blakestad},
  \citenamefont {Britton}, \citenamefont {Jost}, \citenamefont {Knill},
  \citenamefont {Langer}, \citenamefont {Ozeri}, \citenamefont {Seidelin},\
  and\ \citenamefont {Wineland}}]{reichle06}%
  \BibitemOpen
  \bibfield  {author} {\bibinfo {author} {\bibfnamefont {R.}~\bibnamefont
  {Reichle}}, \bibinfo {author} {\bibfnamefont {D.}~\bibnamefont {Leibfried}},
  \bibinfo {author} {\bibfnamefont {R.}~\bibnamefont {Blakestad}}, \bibinfo
  {author} {\bibfnamefont {J.}~\bibnamefont {Britton}}, \bibinfo {author}
  {\bibfnamefont {J.}~\bibnamefont {Jost}}, \bibinfo {author} {\bibfnamefont
  {E.}~\bibnamefont {Knill}}, \bibinfo {author} {\bibfnamefont
  {C.}~\bibnamefont {Langer}}, \bibinfo {author} {\bibfnamefont
  {R.}~\bibnamefont {Ozeri}}, \bibinfo {author} {\bibfnamefont
  {S.}~\bibnamefont {Seidelin}}, \ and\ \bibinfo {author} {\bibfnamefont
  {D.}~\bibnamefont {Wineland}},\ }\href
  {http://onlinelibrary.wiley.com/doi/10.1002/prop.200610326/abstract}
  {\bibfield  {journal} {\bibinfo  {journal} {Fortschritte der Physik}\
  }\textbf {\bibinfo {volume} {54}},\ \bibinfo {pages} {666} (\bibinfo {year}
  {2006})},\ \Eprint {http://arxiv.org/abs/0606237v1} {arXiv:0606237v1
  [arXiv:quant-ph]} \BibitemShut {NoStop}%
\bibitem [{\citenamefont {Hucul}\ \emph {et~al.}(2008)\citenamefont {Hucul},
  \citenamefont {Yeo}, \citenamefont {Olmschenk}, \citenamefont {Monroe},
  \citenamefont {Hensinger},\ and\ \citenamefont {Rabchuk}}]{hucul08}%
  \BibitemOpen
  \bibfield  {author} {\bibinfo {author} {\bibfnamefont {D.}~\bibnamefont
  {Hucul}}, \bibinfo {author} {\bibfnamefont {M.}~\bibnamefont {Yeo}}, \bibinfo
  {author} {\bibfnamefont {S.}~\bibnamefont {Olmschenk}}, \bibinfo {author}
  {\bibfnamefont {C.}~\bibnamefont {Monroe}}, \bibinfo {author} {\bibfnamefont
  {W.~K.}\ \bibnamefont {Hensinger}}, \ and\ \bibinfo {author} {\bibfnamefont
  {J.}~\bibnamefont {Rabchuk}},\ }\href
  {http://dl.acm.org/citation.cfm?id=2016976.2016977} {\bibfield  {journal}
  {\bibinfo  {journal} {Quantum Info. Comput.}\ }\textbf {\bibinfo {volume}
  {8}},\ \bibinfo {pages} {501} (\bibinfo {year} {2008})}\BibitemShut {NoStop}%
\bibitem [{\citenamefont {Torrontegui}\ \emph {et~al.}(2011)\citenamefont
  {Torrontegui}, \citenamefont {Ib\'a\~nez}, \citenamefont {Chen},
  \citenamefont {Ruschhaupt}, \citenamefont {Gu\'ery-Odelin},\ and\
  \citenamefont {Muga}}]{torrontegui11}%
  \BibitemOpen
  \bibfield  {author} {\bibinfo {author} {\bibfnamefont {E.}~\bibnamefont
  {Torrontegui}}, \bibinfo {author} {\bibfnamefont {S.}~\bibnamefont
  {Ib\'a\~nez}}, \bibinfo {author} {\bibfnamefont {X.}~\bibnamefont {Chen}},
  \bibinfo {author} {\bibfnamefont {A.}~\bibnamefont {Ruschhaupt}}, \bibinfo
  {author} {\bibfnamefont {D.}~\bibnamefont {Gu\'ery-Odelin}}, \ and\ \bibinfo
  {author} {\bibfnamefont {J.~G.}\ \bibnamefont {Muga}},\ }\href {\doibase
  10.1103/PhysRevA.83.013415} {\bibfield  {journal} {\bibinfo  {journal} {Phys.
  Rev. A}\ }\textbf {\bibinfo {volume} {83}},\ \bibinfo {pages} {013415}
  (\bibinfo {year} {2011})}\BibitemShut {NoStop}%
\bibitem [{\citenamefont {Dehmelt}(1967)}]{dehmelt67}%
  \BibitemOpen
  \bibfield  {author} {\bibinfo {author} {\bibfnamefont {H.}~\bibnamefont
  {Dehmelt}},\ }\href@noop {} {\bibfield  {journal} {\bibinfo  {journal}
  {Advances in Atomic and Molecular Physics}\ }\textbf {\bibinfo {volume}
  {3}},\ \bibinfo {pages} {53} (\bibinfo {year} {1967})}\BibitemShut {NoStop}%
\bibitem [{\citenamefont {Turner}(1987)}]{turner87}%
  \BibitemOpen
  \bibfield  {author} {\bibinfo {author} {\bibfnamefont {L.}~\bibnamefont
  {Turner}},\ }\href@noop {} {\bibfield  {journal} {\bibinfo  {journal} {Phys.
  Fluids}\ }\textbf {\bibinfo {volume} {30}},\ \bibinfo {pages} {3196}
  (\bibinfo {year} {1987})}\BibitemShut {NoStop}%
\bibitem [{\citenamefont {Drewsen}\ \emph {et~al.}(1998)\citenamefont
  {Drewsen}, \citenamefont {Brodersen}, \citenamefont {Hornek\ae{}r},
  \citenamefont {Hangst},\ and\ \citenamefont {Schiffer}}]{drewsen98}%
  \BibitemOpen
  \bibfield  {author} {\bibinfo {author} {\bibfnamefont {M.}~\bibnamefont
  {Drewsen}}, \bibinfo {author} {\bibfnamefont {C.}~\bibnamefont {Brodersen}},
  \bibinfo {author} {\bibfnamefont {L.}~\bibnamefont {Hornek\ae{}r}}, \bibinfo
  {author} {\bibfnamefont {J.~S.}\ \bibnamefont {Hangst}}, \ and\ \bibinfo
  {author} {\bibfnamefont {J.~P.}\ \bibnamefont {Schiffer}},\ }\href@noop {}
  {\bibfield  {journal} {\bibinfo  {journal} {Phys. Rev. Lett.}\ }\textbf
  {\bibinfo {volume} {81}},\ \bibinfo {pages} {2878} (\bibinfo {year}
  {1998})}\BibitemShut {NoStop}%
\bibitem [{\citenamefont {Nosé}(1984)}]{nose84}%
  \BibitemOpen
  \bibfield  {author} {\bibinfo {author} {\bibfnamefont {S.}~\bibnamefont
  {Nosé}},\ }\href {\doibase 10.1063/1.447334} {\bibfield  {journal} {\bibinfo
  {journal} {The Journal of Chemical Physics}\ }\textbf {\bibinfo {volume}
  {81}},\ \bibinfo {pages} {511} (\bibinfo {year} {1984})}\BibitemShut
  {NoStop}%
\bibitem [{\citenamefont {Pollock}\ and\ \citenamefont
  {Hansen}(1973)}]{pollock73}%
  \BibitemOpen
  \bibfield  {author} {\bibinfo {author} {\bibfnamefont {E.~L.}\ \bibnamefont
  {Pollock}}\ and\ \bibinfo {author} {\bibfnamefont {J.~P.}\ \bibnamefont
  {Hansen}},\ }\href@noop {} {\bibfield  {journal} {\bibinfo  {journal} {Phys.
  Rev. A}\ }\textbf {\bibinfo {volume} {8}},\ \bibinfo {pages} {3110} (\bibinfo
  {year} {1973})}\BibitemShut {NoStop}%
\bibitem [{\citenamefont {Haze}\ \emph {et~al.}(2012)\citenamefont {Haze},
  \citenamefont {Tateishi}, \citenamefont {Noguchi}, \citenamefont {Toyoda},\
  and\ \citenamefont {Urabe}}]{haze2012}%
  \BibitemOpen
  \bibfield  {author} {\bibinfo {author} {\bibfnamefont {S.}~\bibnamefont
  {Haze}}, \bibinfo {author} {\bibfnamefont {Y.}~\bibnamefont {Tateishi}},
  \bibinfo {author} {\bibfnamefont {A.}~\bibnamefont {Noguchi}}, \bibinfo
  {author} {\bibfnamefont {K.}~\bibnamefont {Toyoda}}, \ and\ \bibinfo {author}
  {\bibfnamefont {S.}~\bibnamefont {Urabe}},\ }\href {\doibase
  10.1103/PhysRevA.85.031401} {\bibfield  {journal} {\bibinfo  {journal}
  {Physical Review A}\ }\textbf {\bibinfo {volume} {85}} (\bibinfo {year}
  {2012}),\ 10.1103/PhysRevA.85.031401}\BibitemShut {NoStop}%
\bibitem [{\citenamefont {Blakestad}\ \emph {et~al.}(2011)\citenamefont
  {Blakestad}, \citenamefont {Ospelkaus}, \citenamefont {Vandevender},
  \citenamefont {Wesenberg}, \citenamefont {Biercuk}, \citenamefont
  {Leibfried},\ and\ \citenamefont {Wineland}}]{blakestad11}%
  \BibitemOpen
  \bibfield  {author} {\bibinfo {author} {\bibfnamefont {R.~B.}\ \bibnamefont
  {Blakestad}}, \bibinfo {author} {\bibfnamefont {C.}~\bibnamefont
  {Ospelkaus}}, \bibinfo {author} {\bibfnamefont {A.~P.}\ \bibnamefont
  {Vandevender}}, \bibinfo {author} {\bibfnamefont {J.~H.}\ \bibnamefont
  {Wesenberg}}, \bibinfo {author} {\bibfnamefont {M.~J.}\ \bibnamefont
  {Biercuk}}, \bibinfo {author} {\bibfnamefont {D.}~\bibnamefont {Leibfried}},
  \ and\ \bibinfo {author} {\bibfnamefont {D.~J.}\ \bibnamefont {Wineland}},\
  }\href {\doibase 10.1103/PhysRevA.84.032314} {\bibfield  {journal} {\bibinfo
  {journal} {Phys. Rev. A}\ }\textbf {\bibinfo {volume} {032314}},\ \bibinfo
  {pages} {1} (\bibinfo {year} {2011})}\BibitemShut {NoStop}%
\bibitem [{\citenamefont {Schulz}\ \emph {et~al.}(2006)\citenamefont {Schulz},
  \citenamefont {Poschinger}, \citenamefont {Singer},\ and\ \citenamefont
  {Schmidt-Kaler}}]{schulz06}%
  \BibitemOpen
  \bibfield  {author} {\bibinfo {author} {\bibfnamefont {S.}~\bibnamefont
  {Schulz}}, \bibinfo {author} {\bibfnamefont {U.}~\bibnamefont {Poschinger}},
  \bibinfo {author} {\bibfnamefont {K.}~\bibnamefont {Singer}}, \ and\ \bibinfo
  {author} {\bibfnamefont {F.}~\bibnamefont {Schmidt-Kaler}},\ }\href {\doibase
  10.1002/prop.200610324} {\bibfield  {journal} {\bibinfo  {journal}
  {Fortschritte der Physik}\ }\textbf {\bibinfo {volume} {54}},\ \bibinfo
  {pages} {648} (\bibinfo {year} {2006})}\BibitemShut {NoStop}%
\bibitem [{\citenamefont {Singer}\ \emph {et~al.}(2010)\citenamefont {Singer},
  \citenamefont {Poschinger}, \citenamefont {Murphy}, \citenamefont {Ivanov},
  \citenamefont {Ziesel}, \citenamefont {Calarco},\ and\ \citenamefont
  {Schmidt-{K}aler}}]{singer10}%
  \BibitemOpen
  \bibfield  {author} {\bibinfo {author} {\bibfnamefont {K.}~\bibnamefont
  {Singer}}, \bibinfo {author} {\bibfnamefont {U.}~\bibnamefont {Poschinger}},
  \bibinfo {author} {\bibfnamefont {M.}~\bibnamefont {Murphy}}, \bibinfo
  {author} {\bibfnamefont {P.}~\bibnamefont {Ivanov}}, \bibinfo {author}
  {\bibfnamefont {F.}~\bibnamefont {Ziesel}}, \bibinfo {author} {\bibfnamefont
  {T.}~\bibnamefont {Calarco}}, \ and\ \bibinfo {author} {\bibfnamefont
  {F.}~\bibnamefont {Schmidt-{K}aler}},\ }\href {\doibase
  10.1103/RevModPhys.82.2609} {\bibfield  {journal} {\bibinfo  {journal} {Rev.
  Mod. Phys.}\ }\textbf {\bibinfo {volume} {82}},\ \bibinfo {pages} {2609}
  (\bibinfo {year} {2010})}\BibitemShut {NoStop}%
\bibitem [{sim()}]{simion}%
  \BibitemOpen
  \href@noop {} {}\bibinfo {note} {Http://www.simion.com}\BibitemShut {NoStop}%
\bibitem [{\citenamefont {Alheit}\ \emph {et~al.}(1995)\citenamefont {Alheit},
  \citenamefont {Henning}, \citenamefont {Morgenstern}, \citenamefont {Vedel},\
  and\ \citenamefont {Werth}}]{alheit95}%
  \BibitemOpen
  \bibfield  {author} {\bibinfo {author} {\bibfnamefont {R.}~\bibnamefont
  {Alheit}}, \bibinfo {author} {\bibfnamefont {C.}~\bibnamefont {Henning}},
  \bibinfo {author} {\bibfnamefont {R.}~\bibnamefont {Morgenstern}}, \bibinfo
  {author} {\bibfnamefont {F.}~\bibnamefont {Vedel}}, \ and\ \bibinfo {author}
  {\bibfnamefont {G.}~\bibnamefont {Werth}},\ }\href@noop {} {\bibfield
  {journal} {\bibinfo  {journal} {Appl. Phys. B}\ }\textbf {\bibinfo {volume}
  {61}},\ \bibinfo {pages} {277} (\bibinfo {year} {1995})}\BibitemShut
  {NoStop}%
\end{thebibliography}
%

\end{document}